\documentstyle[12pt]{article}
\begin{document}
\input{epsf}
\def\epp{\epsilon^{\prime}}
\def\vep{\varepsilon}
\def\ra{\rightarrow}
\def\ppg{\pi^+\pi^-\gamma}
\def\vp{{\bf p}}
\def\ko{K^0}
\def\kb{\bar{K^0}}
\def\al{\alpha}
\def\ab{\bar{\alpha}}
\def\be{\begin{equation}}
\def\ee{\end{equation}}
\def\bea{\begin{eqnarray}}
\def\eea{\end{eqnarray}}
\def\question#1{{{\marginpar{\small \sc #1}}}}
\def\mpl{{{m_{Pl}}}}
\def\oph{{{\Omega_{\widetilde{\gamma}}h^2}}}
\def\be{\begin{equation}}
\def\ee{\end{equation}}
\def\ba{\begin{eqnarray}}
\def\ea{\end{eqnarray}}
\def\la{\mathrel{\mathpalette\fun <}}
\def\ga{\mathrel{\mathpalette\fun >}}
\def\fun#1#2{\lower3.6pt\vbox{\baselineskip0pt\lineskip.9pt
        \ialign{$\mathsurround=0pt#1\hfill##\hfil$\crcr#2\crcr\sim\crcr}}}
\def\pho{{{\widetilde{\gamma}}}}
\def\r0{{{R^0}}}
\def\glu{{{\widetilde{g}}}}
\def\sec{{{\mbox{sec}}}}
\def\GeV{{{\mbox{GeV}}}}
\def\MeV{{{\mbox{MeV}}}}
\def\SUSY{{{{\sc susy}}}}
\def\LSP{{{{\sc lsp}}}}
\def\LEP{{{{\sc lep}}}}
\def\LROCS{{{{\sc lrocs}}}}
\def\WIMPS{{{{\sc wimps}}}}
\def\cm{{{\mbox{cm}}}}
\def\photino{{{\mbox{photino}}}}
\def\gluino{{{\mbox{gluino}}}}
\def\mb{{{\mbox{mb}}}}
\def\avg#1{{{{\langle #1 \rangle }}}}
\def\taun{{{\tau_{9}}}}
\def\mpl{{{m_{Pl}}}}
\def\re#1{{[\ref{#1}]}}
\def\eqr#1{{Eq.\ (\ref{#1})}}
\def\mst{{{M_{\widetilde{S}}}}}
\newcommand{\gl}{\tilde{g}}
\newcommand{\sneu}{\tilde{\nu}}
\newcommand{\sq}{\tilde{q}}
\newcommand{\se}{\tilde{e}}
\newcommand{\ch}{\chi^{\pm}}
\newcommand{\neut}{\chi^{0}}
\newcommand{\gsi}{\,\raisebox{-0.13cm}{$\stackrel{\textstyle>}
{\textstyle\sim}$}\,}
\newcommand{\lsi}{\,\raisebox{-0.13cm}{$\stackrel{\textstyle<}
{\textstyle\sim}$}\,}

\rightline{RU-97-102}  
\baselineskip=18pt
\vskip 0.7in
\begin{center}
{\bf \LARGE Can Ultra High Energy Cosmic Rays\\ be Evidence for New Particle
Physics?}\\
\vspace*{0.9in}
{\large Glennys R. Farrar}\footnote{Invited talk at the Workshop on "Observing
the Highest Energy Particles ($> 10^{20}$ eV) from Space", College Park,
MD, Nov. 13-15, 1997.  Research supported in part by NSF-PHY-94-2302.}\\ 
{\it Department of Physics and Astronomy}\\
{\it Rutgers University, Piscataway, NJ 08855, USA}\\
\vspace*{0.1in} 
\end{center}

\vskip  0.5in

{\bf Abstract:}
Candidate astrophysical acceleration sites capable of producing the highest
energy cosmic rays ($E > 10^{19.5}$ eV) appear to be at far greater distances
than is compatible with their being known particles.  The
properties of a new particle which can account for observations are discussed
and found to be tightly constrained.  In order to travel 100's or 1000's of Mpc
through the cosmic microwave background radiation without severe energy
loss and yet produce a shower in Earth's atmosphere which is
consistent with observations, it must be a hadron with mass of order a few
GeV and lifetime greater than about 1 week.  A particle with the required
properties was identified years ago in the context of supersymmetric theories
with a very light gluino.  Laboratory experiments do not exclude it, as is
discussed briefly. 

\thispagestyle{empty}
\newpage
\addtocounter{page}{-1}

\section*{Introduction}
\hspace*{2em}
Properties of ultra-high-energy cosmic rays (UHECRs) present a paradox
within the standard astrophysical and particle physics framework.  According
to the Greisen-Zatsepin-Kuzmin bound\cite{gzk}, the spectrum of cosmic
rays should cut off at about $10^{19.5}$ eV unless the source is closer than
of order tens of megaparsecs.  Yet many cosmic ray events have been
observed at higher energy than this, and suitable cosmic accelerators
within the GKZ range have not been found\cite{akeno,flyseye,ElbSom,biermann:rev}.
Some have considered the possibility of exotic relatively nearby sources,
e.g., decay of super long-lived relics of the big bang\cite{kr,berezinsky:cr}
or associated production in gamma ray bursts\cite{waxman:cr}.  Another possibility
is that the UHECRs are exotic particles which can be transmitted from cosmological
distances of order 100-1000's Mpc, where suitable conventional accelerators
are found\cite{f:104,f:114}.  It turns out that it is difficult for a particle
to simultaneously have the properties necessary to evade the GZK bound and
also interact like an ordinary hadron in the atmosphere, as the UHECRs
do\cite{f:114}.  However, somewhat miraculously, a particle with the correct
properties to account for the UHECRs is automatically present in an interesting
class of supersymmetric theories.  The properties of this particle were
delineated before the observation of the UHECRs and are not subject to
extensive tuning\cite{f:51,f:104}.

\section*{Brief review of light gluino phenomenology}
\hspace*{2em}
A supersymmetric version of a locally gauge invariant theory such as
the standard model necessarily has massless fermions -- the superpartners
of the gauge bosons -- called gauginos.  Therefore the theory
has an ``accidental''\footnote{A global symmetry which is not imposed by hand
but is the inevitable consequence of the gauge symmetries of the theory,
like baryon and lepton number in the standard model, is called an accidental
symmetry.} chiral symmetry which I will generically call $R$-invariance
below.  When the gauge invariance and supersymmetry are spontaneously broken,
the $R$-invariance may or may not break.  

In supersymmetry (SUSY) breaking scenarios which do not break $R$-invariance
at all, every gaugino is massless at tree level\footnote{In a typical
supergravity case, all dimension-3 SUSY breaking terms are absent so that
in conventional notation, $m_1,~m_2,~m_3,~A \approx 0$.} and $R$-parity is
conserved.  Such SUSY breaking has several attractive theoretical consequences
such as the absence of the ``SUSY CP problem"\cite{f99101,f:104}. Gauginos
get calculable masses through radiative corrections from electroweak
(gaugino/higgsino-Higgs/gauge 
boson) and top-stop loops\cite{radmass}.  Evaluating these loops within the
allowed parameter space leads to a gluino mass range $m_{\tilde{g}}\sim
\frac{1}{10} - \frac{1}{2}$ GeV\cite{f99101,f:104}, while analysis of the
$\eta'$ mass and properties narrows this to $m(\glu) \approx 120 \pm 40$
MeV\cite{f:108}. The photino mass range depends on more unknowns than the
gluino mass, such as the higgs and higgsino sectors, but can be estimated
to be $m_{\tilde{\gamma}} \sim \frac{1}{10} - 1 \frac{1}{2}$ GeV\cite{f99101}. 

In other SUSY breaking scenarios some or all gauginos are massive, with
masses of order the squark or slepton masses.  For purposes of understanding
ultra-high-energy cosmic rays, we will be interested in models in which
at least the gluino is massless or light at tree level.  Such models have
been constructed in the gauge-mediated framework in refs.
\cite{mohapetal,raby}.  In Raby's model the gluino mass can be tuned 
over a range $\sim 100$ GeV, and the gluino can be the lightest supersymmetric
particle. 

If the gluino lifetime is long compared with the strong interaction time
scale, $10^{-23}$ sec, it binds with quarks, antiquarks and/or gluons to
make color-singlet hadrons (generically called $R$-hadrons\cite{f2324}). 
The lightest of these is expected to be the gluino-gluon bound state,
designated $\r0$.  It is predicted to have a mass in the range
$1.3-2.2$ GeV\cite{f:95,f:104}, approximately degenerate with the
lightest glueball ($0^{++}$) and ``gluinoball'' ($0^{-+},~\glu
\glu$), for a gluino mass of $\approx 100$ MeV\cite{f99101,f:108}.
The existance of an ``extra'' isosinglet pseudoscalar meson, $\eta(1410)$,
which is difficult to accomodate in standard QCD but which matches nicely
the mass and properties predicted for the pseudoscalar $\glu \glu$, is
encouraging for the light gluino ansatz\cite{f:109,f:104}.  

The lightest $R$-hadron with non-zero baryon number is the $ uds\glu$
bound state designated $S^0$\cite{f:51}.  As we shall see in the next
section, the very highest energy cosmic rays reaching Earth\cite{f:104} could
well be $S^0$'s.  On account of the very strong hyperfine attraction among the
quarks in the flavor-singlet channel, the $S^0$ mass is about $210 \pm 20$
MeV lower than that of the lightest $R$-nucleons\cite{f:51,f:52}.  If we
knew the mass of the crypto-exotic flavor singlet baryon $uds$-gluon, we
could place the $S^0$ mass to within a couple of hundred MeV by
analogy\footnote{Due to the heavy mass of squarks, no rigorous supersymmetry
argument equates the $uds\glu$ and $udsg$ masses in the massless gluino limit,
unlike the case for glueball and glueballino in quenched approximation.}. 
The $1/2^{-}$ baryon $\Lambda(1405)$ could be this $udsg$
state\cite{f:104}, in which case we see the $S^0$ mass could be as low as
about 1.5 GeV.  We can obtain an estimate of the maximum possible mass
of the $uds\glu$ as follows.  Add the mass of a $\Lambda$ (to account
for the effective mass of the quarks in a color singlet state, in the presence
of chiral symmetry breaking) to the mass of a glueball (to account for the
mass associated with combining a pair of color octets whose short-distance
mass is zero) plus a bare gluino mass of 120 MeV: $m(S^0) < 1120 + 1600 +
120 = 2740$ MeV.  This estimate does not account for the hyperfine attraction
due to the $uds$ in an $S^0$ being in a flavor singlet state whereas the
$uds$ in a $\Lambda$ is in a flavor octet state\cite{f:51,f:52}.  A more
realistic estimate would replace the $\Lambda$ mass of 1120 MeV with the
nucleon mass minus 210 MeV, leading to $m(S^0) \approx 940 - 210 + 1600 +
120 = 2450$ MeV. 

If new gluino-containing hadrons have lifetimes shorter than about $10^{-10}$
sec, they can be discovered through missing energy or beam dump
experiments\cite{f2324}.  However if the gluino is nearly massless it
is long enough lived that the standard techniques are inapplicable\cite{f:51}.
The non-negligible radiative mass for the photino compared to the $\r0$,
leads to an $\r0$ lifetime in the range $10^{-10} - 10^{-5}$ sec\cite{f99101,f:104}
if the photino mass is radiative.  The dominant decay mode is $\r0 \rightarrow
\pi^+ \pi^- \pho$\cite{f:104}.  If neutralinos have tree level masses
large compared to the gluino's, the $R^0$ would be stable or very long-lived,
depending on the mass of the gravitino.  When the $R^0$ is stable or long-lived,
the usual SUSY 
signatures relying on prompt neutralino or goldstino (gravitino)
production\cite{f2324} are not useful\cite{f:51}.  As a consequence, gluino
masses less than about $ \frac{1}{2}$ GeV are largely unconstrained\cite{f:95}\footnote{The
ALPEH claim to exclude light gluinos\cite{aleph:lg} assigns a $1 \sigma$
theoretical systematic error based on varying the renormalization scale over
a small range. Taking a more generally accepted range of scale variation
and accounting for the large sensitivity to hadronization model, 
the ALEPH systematic uncertainty is comparable to that of other
experiments and does not exclude light gluinos\cite{fLaT,f:119}.  The claim
of Nagy and Troscsanyi\cite{nt:lg}, that use of $R_4$ allows a 95\% cl exclusion,
has even worse problems.  In addition to scale sensitivity, their result relies
on using the central value of $\alpha_s$.  When the error bars on $\alpha_s$
are included, their limit is reduced to $1 \sigma$, even without
considering the uncertainty due to scale and resummation scheme sensitivity.}.
Proposals for direct searches for hadrons containing gluinos, via
their decays in $K^0$ beams and otherwise, are given in Refs.
\cite{f:95,f:104}.  For the moment, the experimental cuts preclude investigating
the parameter ranges of theoretical interest, but part of the parameter
space relevant when photinos provide dark matter should be amenable to
study\footnote{See \cite{ktev:lg} for a first experimental effort at
placing a limit on $\r0$ production and decay via $\pi^+ \pi^- \pho$.}
In the course of the next two years it should be possible to exclude the
all-gauginos-light scenario if it is not correct\cite{f:116}.  For a recent
detailed survey of the experimental constraints on light gaugino scenarios,
see \cite{f:119}.  

An attractive feature of models with all gauginos massless or extremely light
at tree level is that relic photinos naturally provide the correct abundance
of dark matter\cite{f:100,f:113}.  By contrast, finding a good explanation
for dark matter is a problem if the gluino is the only light gaugino.  In
this case the lightest neutralino would not be stable on cosmological time 
scales and thus could not be the dark matter particle.  The gravitino is
not a satisfactory dark matter candidate, if that is the lightest susymmetric
particle, due to structure formation considerations\cite{masiero:dm}.
Nor can the lightest gluino-containing hadrons provide sufficient relic dark
matter density, even if absolutely stable as in the model of \cite{raby},
because they annihilate too efficiently\cite{plaga,f:113}\footnote{In
order to produce an interesting dark matter density, the gluino mass must
be so large that it is inconsistent with properties of our
galaxy\cite{nardi_roulet}.}.  However new types of matter with conserved
quantum numbers can be present in the theory, so the absence of a neutralino
dark matter candidate may not be an insurmountable problem in models in
which there is a light gluino but no light photino.  

Now let us turn to the lifetime of the $S^0$, recalling the relevant mass
estimates above for a gluino mass of 120 MeV: $m(\r0) = 1.3-2.2$ GeV, $m(S^0)
= 1.4 - 2.7$ GeV, and $m_{\tilde{\gamma}}$ must lie in the range $ \sim 0.9
- 1.7$ GeV if photinos account for the relic dark matter\cite{f:113}.  Thus
the strong-interaction decay $S^0 \rightarrow \Lambda \r0$ is unlikely to
be kinematically allowed, nor does the weak-interaction decay $S^0 \rightarrow
n \r0$ seem likely.  Even if kinematically allowed, the decay $S^0 \rightarrow
n \pho$ would lead to a long $S^0$ lifetime since it involves a
flavor-changing-neutral-weak transition mediated by squarks.  However the
$S^0$ may be kinematically unable to decay since the mass estimates above
are compatible with $m(S^0) < m(p) + m(e^-) + m_{\tilde{\gamma}}$\cite{f:104}.
Requiring the $S^0$ to be stable or very long lived leads to the favored
mass range $1.5 \la m(S^0) \la  2.6$ GeV.  

If a gluino-containing hadron is absolutely stable, the most important consideration
is whether it binds to nucleons to produce new stable nuclei which accumulate
near Earth\cite{f:51}.  If so, limits on exotic isotopes give stringent
limits on their abundance\cite{f:51,f:95,plaga,nardi_roulet,nussinov:lg}.
The $S^0$ is not expected to bind to nuclei\cite{f:95}.  The large ($\gsi
400$ MeV) energy gap to intermediate states accessible by pion exchange implies
the effective nuclear potential seen by an $S^0$ is too shallow to
support a bound state, except conceivably for very heavy nuclei\footnote{See
\cite{nussinov:lg} for an approach to estimating the nuclear-size dependence
of the effective potential.  It is less clear that a stable $R^0$ or $u
\bar{d} \gl$ would not bind to nuclei and therefore be excluded;  see
\cite{plaga,raby,nussinov:lg} for a discussion of some of the issues.}. 

To summarize, the gluino may be extremely light ($m_{\glu} \approx 120$
MeV) and give rise to new hadrons with masses below 3 GeV.  Such a
scenario predicts a particle like the $\eta(1410)$ and thus resolves the
mystery of the existance of this state.  As we shall see in the next
section, the lightest $R$-baryon can naturally account for the
ultra-high-energy cosmic rays which have been 
observed above the GKZ bound.  The other gauginos may also be light.
Such scenarios are very attractive because the photino naturally
accounts for relic dark matter and there is no SUSY-CP problem.  The
all-gauginos-light scenario should be excludable within a year or so via
LEP experiments\cite{f:116}.  The only-gluino-light scenario will be more
difficult to exclude because that will require better theoretical control
of perturbative and non-perturbative effects in QCD\cite{f:119}. 

\section*{Ultra-High-Energy Cosmic Rays}
\hspace*{2em}
If the light gaugino scenario is correct, the lightest $R$-baryon,
$S^0 \equiv uds\glu$\cite{f:51}, may be responsible for the very highest
energy cosmic rays reaching Earth\cite{f:104}.  As is well-known to this audience,
the observation of several events with energies $\ga 2~ 10^{20}$ eV\cite{akeno,flyseye}
presents a severe puzzle for astrophysics\footnote{For an introduction and
references see \cite{uhecr} and talks at this conference.}.  Protons with
such high energies have a large scattering cross section on the cosmic microwave
background photons, because $E_{cm}$ can be sufficient to excite the
$\Delta(1230)$ resonance\cite{gzk}.  Consequently the scattering length
of ultra-high-energy protons is of order 10 Mpc or less.  The upper bound
on the energy of cosmic rays which could have originated in the local cluster,
$\sim 10^{19.5}$ eV, is called the Greisen-Zatsepin-Kuzmin (GZK) bound.  

\begin{figure}[p]
\hspace*{25pt} \epsfxsize=400pt \epsfbox{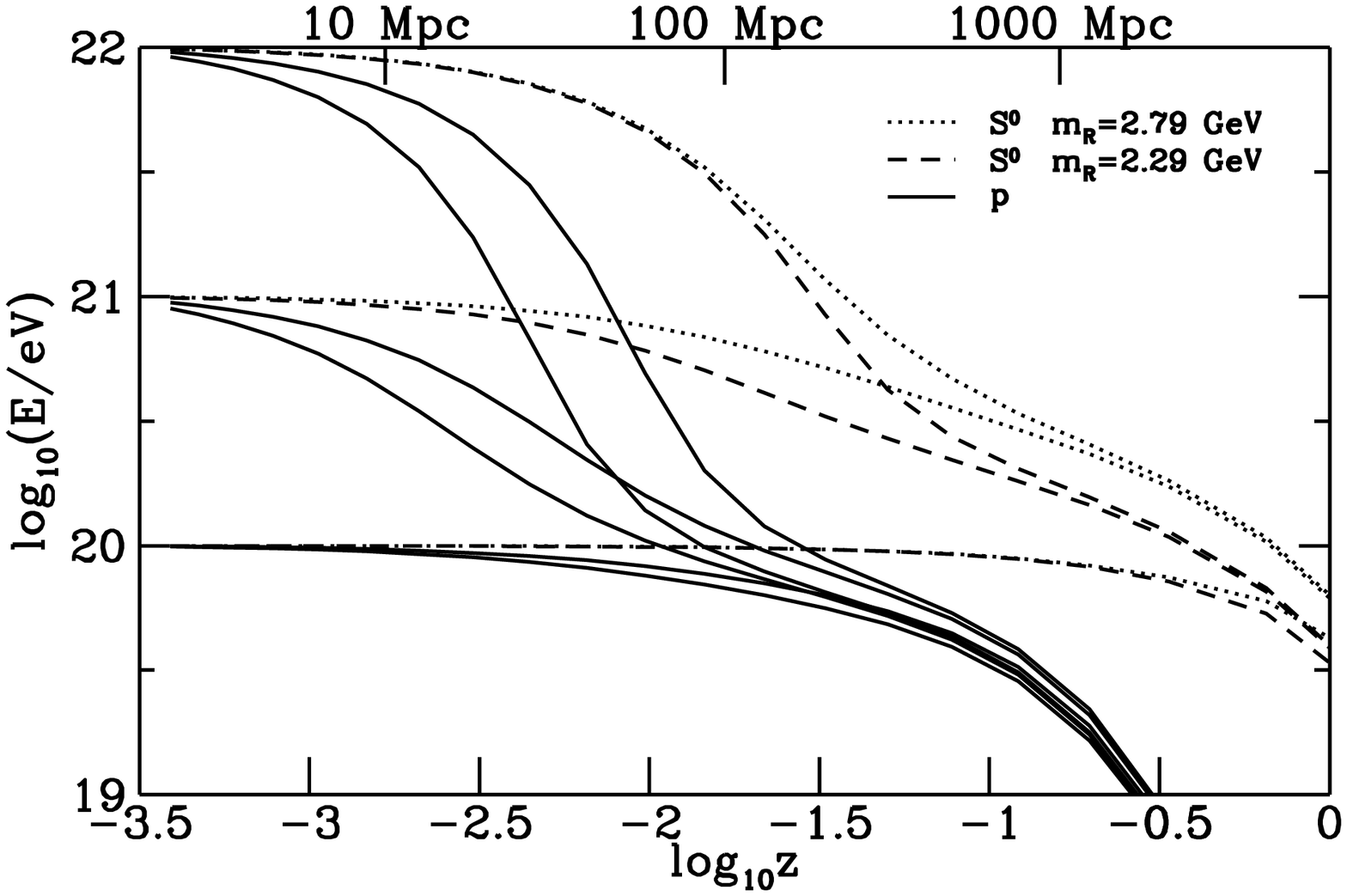}\\
\hspace*{25pt} \epsfxsize=400pt \epsfbox{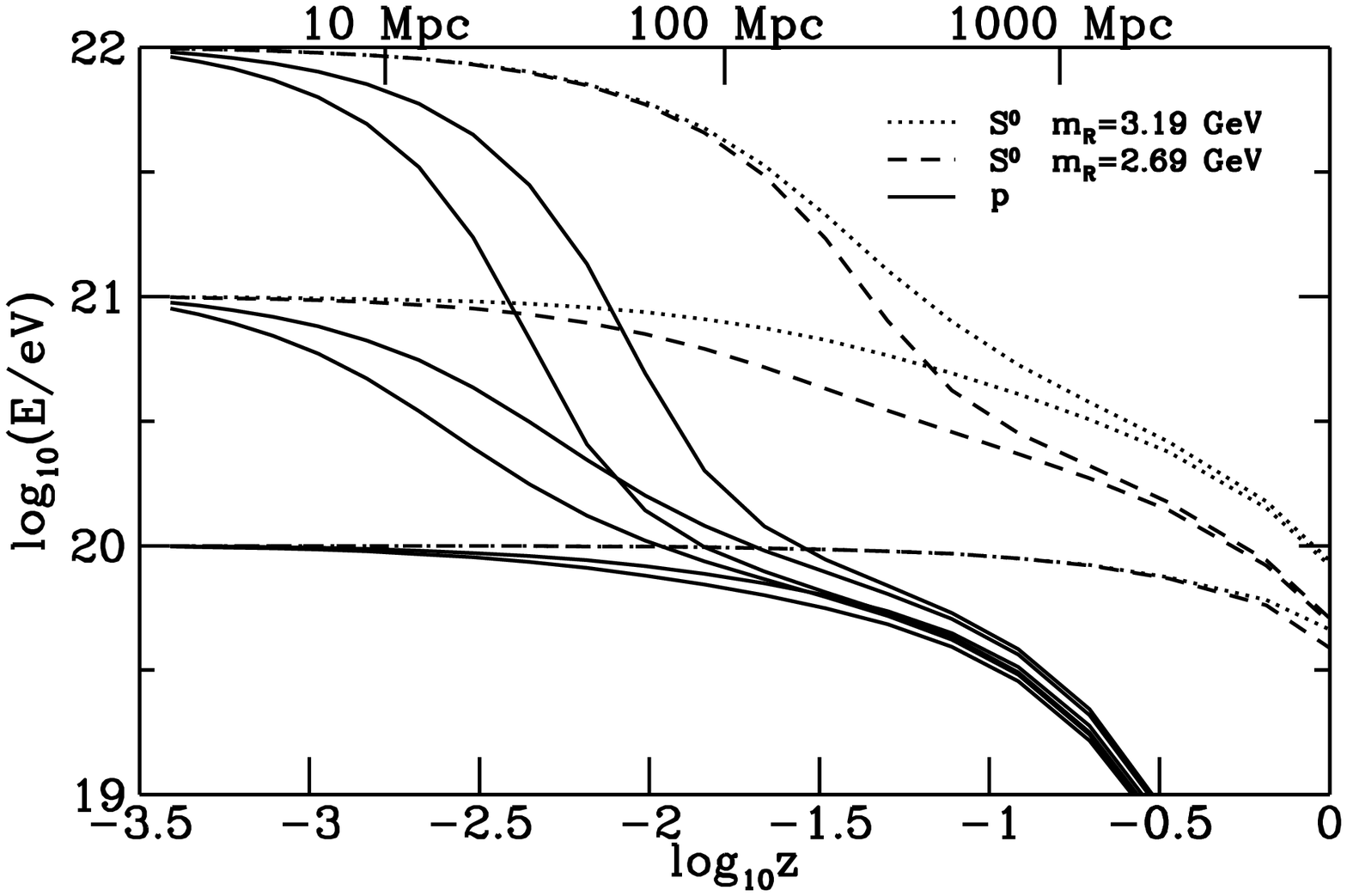}\\
\caption[]{ The figures show the
primary particle's energy as it would be observed on Earth today if it
were injected with various energies ($10^{22}eV$ eV, $10^{21}$ eV, and
$10^{20}$ eV) at various redshifts.  The distances correspond to
luminosity distances.  The mass of $S^0$ is $1.9 $GeV in the upper plot
while it is $2.3 $GeV in the lower plot.  
 Here, the Hubble constant has been set to 50
km sec$^{-1}$ Mpc$^{-1}$. }
\label{gzk}
\end{figure}

Two of the highest energy cosmic ray events come from the same
direction in the sky\cite{akeno,flyseye}; the geometrical random probability
for this is $\sim 10^{-3}$ using a 1-sigma error box\footnote{A third event above
the GZK bound in the same direction has also been identified, as well as
another triplet and two other pairs of UHE events\cite{akeno}.}.
The nearest plausible source in that direction is the Seyfert galaxy MCG
8-11-11 (aka UGC 03374), but it is 62-124 Mpc away\cite{ElbSom}.  An even
more attractive source is the AGN 3C 147, but its distance is at least
1200 Mpc\cite{ElbSom}.  The solid curves in Fig. \ref{gzk}, reproduced from
ref. \cite{f:114}, show the spectrum of high energy protons as a function
of their initial distance, for several different values of the injection
energy.  Compton scattering and photoproduction, as well as redshift effects,
have been included\cite{f:114}.  It is evidently highly unlikely that 
the highest energy cosmic ray events can be due to protons from MCG 8-11-11,
and even more unlikely that two or three high energy protons could penetrate
such distances or originate from 3C 147.

It is also unlikely that the UHECR primaries are photons.  First of all,
photons of these energies have a scattering length, 6.6 Mpc, comparable to that of
protons when account is taken of scattering from radio as well
as CMBR photons\cite{ElbSom}.  Secondly, the atmospheric showers appear
to be hadronic rather than electromagnetic: the UHECRs observed via extensive
air shower detectors have the large muon content characteristic of a hadronic
primary and the shower development of the $3.2 \times 10^{20}$ eV Fly's Eye
event has been found to be incompatible with that of a photon primary\cite{halzen}.

However the ground-state $R$-baryon, the flavor singlet scalar $uds
\tilde{g}$ bound state denoted $S^0$, could explain these
ultra-high-energy events\cite{f:104,f:114}.  The $S^0$ lifetime is plausibly
longer than $\sim 10^5$ sec, the proper time required for a few $10^{20}$
GeV particle of mass $ 2$ GeV to travel $ 100$ Mpc.  Furthermore, the GZK
bound for the $S^0$ is several times higher than for protons.  Three effects
contribute to this: (a) The $S^0$ is neutral, so its interactions with photons
cancel at leading order and are only present due to inhomogeneities in its
quark substructure. (b) The $S^0$ is heavier than the proton.  (c) The mass
splitting between the $S^0$ and the lowest lying resonances which can be
reached in a $\gamma S^0$ collision (mass $\equiv M^*$) is larger than the 
proton-$\Delta(1230)$ splitting.  

The threshold energy for exciting the resonances in $\gamma S^0$
collisions is larger than in $\gamma p$ collisions by the factor\cite{f:104}
$\frac{m_{S^0}}{m_p} \frac{( M^* - M_{S^0})}{(M_\Delta - m_p)}$, where
$M^* - M_{S^0}$ is the mass splitting between the $S^0$ and the lowest mass
resonance excited in $\gamma S^0$ scattering.  Since the photon couples as
a flavor octet, the resonances excited in $S^0 \gamma$ collisions are flavor
octets.  Since the $S^0$ has spin-0 and the photon has helicity $\pm
1$, only a spin-1 $R_{\Lambda}$ or $R_{\Sigma}$ can be produced in the
intermediate state.  There are two $R$-baryon flavor octets with
$J=1$, one with total quark spin 3/2 and the other with total quark
spin 1/2, like the $S^0$.  Neglecting the mixing between these states
which is small, their masses are about 385-460 and 815-890 MeV heavier
than the $S^0$, respectively\cite{f:52}.  This is a much larger
splitting than $M_\Delta - m_p = 290~ {\rm MeV}$ Thus one expects that
the GZK bound is a factor of 2.7 - 7.5 higher for $S^0$'s than for
$p$'s, depending on which $R$-hyperons are strongly coupled to the
$\gamma S^0$ system\cite{f:104}.  A detailed calculation of $S^0$
scattering on microwave photons, including $e^+ e^-$ pair production
and redshift effects can be found in \cite{f:114}. The results for a
typical choice of parameters are shown in Fig. \ref{gzk}, confirming
the rough estimate of ref. \cite{f:104}.    

However as the above discussion makes clear, any neutral stable hadron with
mass larger than a few times the proton mass will have a long enough
mean free path in the CMBR to evade the GZK bound.  Many extensions of the
standard model contain stable colored particles besides quarks and gluons
which, due to confinement, will be "clothed" with quarks and gluons to form
new stable hadrons.  However as pointed out in \cite{f:114}, there is also
an upper bound on the mass of an acceptable UHECR primary (uhecron).  This
comes about because the fractional energy loss per collision with atmospheric
nuclei is of order (1 GeV)$/m_U$, where $m_U$ is the mass of the uhecron.
But if the energy loss per collision is too small, the uhecron shower
development does not resemble that of a nucleon. Detailed  Monte Carlo simulation
is necessary to pin down the maximum acceptable mass\cite{afk}, but it seems
unlikely to exceed of order ten GeV.  Therefore new heavy colored particles
whose masses are $\ga 100$ GeV could not be the UHECR primaries even if they
were not excluded otherwise.  Supersymmetry breaking schemes such as \cite{raby}
which allow parameters to be adjusted to make the gluino stable but do not
require it to be nearly massless must be fine tuned to account for the observed
UHECR showers.  It is remarkable that the mass of the $S^0$ and its excitations
in the nearly-massless gluino scenario fortuitously falls in the rather
narrow range required to explain the UHECR's and yet be consistent with present
laboratory and astrophysical constraints.

The question of production/acceleration of UHECR's is a difficult one, even
if the UHECR primary could be a proton.  Mechanisms for accelerating
protons are reviewed in ref. \cite{biermann:rev}.  Most of the mechanisms
proposed for protons have variants which work for $S^0$'s\cite{f:114}.  Indirect
production via decay of defects or long-lived relics of the big bang proceeds
by production of extremely high energy quarks (or gluinos).  Since all baryons
and $R$-baryons eventually decay to protons and $S^0$'s respectively, the
relative probability that a quark or gluino fragments into an $S^0$ compared
to a proton can be expected to be of order $10^{-1}-10^{-2}$.  This estimate
incorporates the difficulty of forming hadrons with increasingly large numbers
of constituents, as reflected in the baryon to meson ratio in quark fragmentation
which is typically of order 1:10.  To be conservative, an additional
possible suppression of up to a factor of 10 is included because the typical
mass of $R$-baryons is greater than that of baryons.

\begin{figure}
\vspace{9pt}
\epsfxsize = 400pt \epsffile{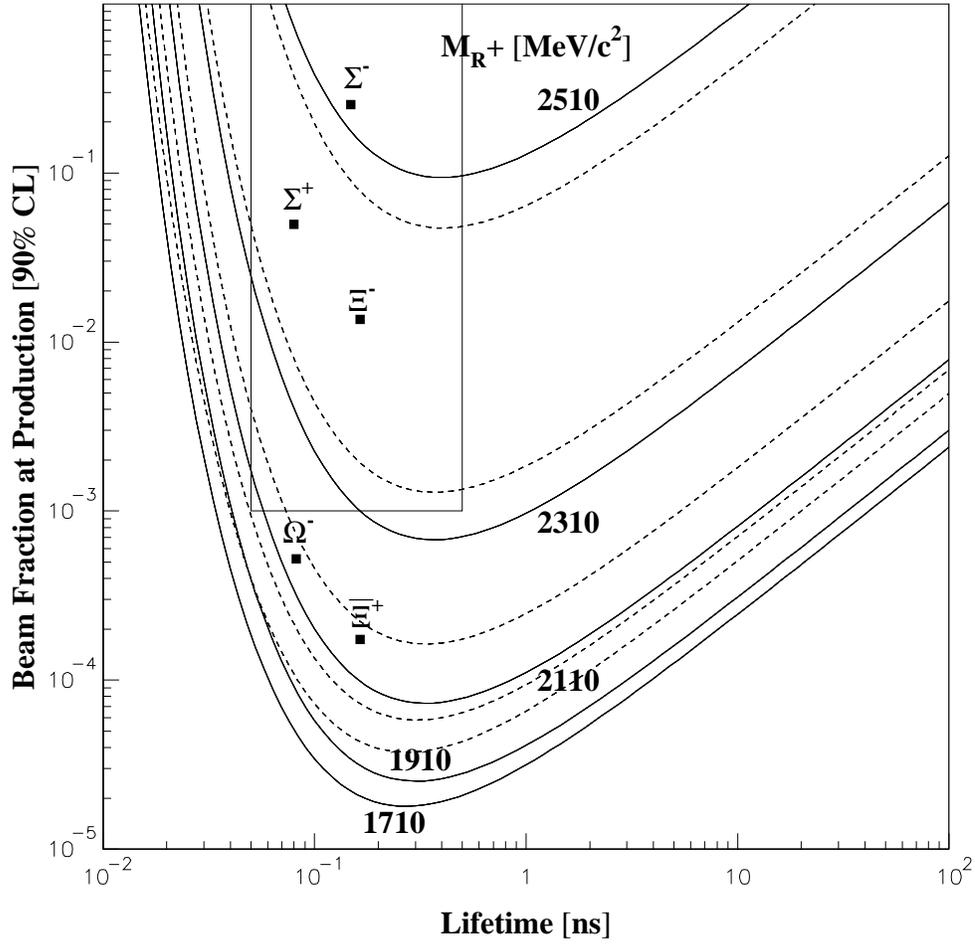}
\caption[]{E761 limits vs. $\tau(R_p)$\cite{e761}.  Solid contours give limits
for various values of the $R_p$ mass, which is about 210 MeV above
the $S^0$ mass.  Ignore the box and dotted contours.}  
\label{e761}
\end{figure}

Mechanisms which accelerate protons also produce high energy $S^0$'s,
via the production of $R_p$'s ($uud \gl$ bound states) in $pN$
collisions\cite{f:114}.  A problem with some proton acceleration mechanisms
which is overcome with $S^0$'s is that astrophysical accelerators capable of producing
ultra-high-energy protons may have such large densities that the protons
are unlikely to escape without colliding and losing energy.  In the scenario
at hand, a high proton collision rate is actually advantageous for producing
$R_p$'s.  These $R_p$'s decay to $S^0 \pi^+$ via a weak interaction, with lifetime
estimated to be $2~\cdot 10^{-11} - 2~\cdot 10^{-10}$ sec\cite{f:104}.
The $S^0 N$ cross section is likely to be smaller than the $N N$ cross section
by up to a factor of 10\cite{f:104}.  Furthermore the $S^0$ interaction with
electrons and photons is negligible.

It may be significant that the predicted time-dilated lifetime of an $R_p$
of energy $ \approx 3~ 10^{20}$ eV, is of order seconds -- a characteristic timescale
for Gamma Ray Bursts.  Mechanisms for producing ultra-high-energy protons in GRB's
would translate to the production of $R_p$'s\cite{waxman:cr}.

Laboratory experiments can be used to get upper bounds on the production
of $R_p$'s, which may be helpful in deciding whether the $S^0$ production
mechanisms discussed above is plausible.  The E761 collaboration at Fermilab searched
for evidence of $R_p \rightarrow S^0 \pi^+$\cite{e761}.  Their result is
shown in Fig. \ref{e761}. If the lifetime of the $R_p$ is of order
nanoseconds and $m(S^0) \la 2.1$ GeV (so $m(R_p) \la 2.3$ GeV),
these limits would make it difficult to produce sufficient high energy $S^0$'s
via $R_p$'s. But for a lifetime of order $2~ 10^{-1} - 2~ 10^{-2}$ ns as
estimated in \cite{f:104} and the favored $S^0$ mass of $\approx 2.4$ GeV,
the E761 limits are too weak to be a constraint. 
As detailed in \cite{f:119}, a second generation experiment of this type
would be very valuable.

\section*{Summary}

Cosmic ray events with energies above the GZK bound may be due to a
quasi-stable hadron containing a very light gluino, such as the $uds\gl$
bound state called $S^0$.  When the gluino mass arises only radiatively,
due to the spontaneous breaking of electroweak symmetry, its mass is about
100 MeV.  This implies a favored $S^0$ mass range of $1.5 - 2.6$ GeV.  The
GZK cutoff for $S^0$'s occurs at higher energy than for protons, and $S^0$'s
of $3 \times 10^{20}$ eV can come from cosmological distances where appropriate
accelerators are found.  The atmospheric shower of a high energy $S^0$ is
similar to that of a nucleon.  

The $S^0$'s are not deflected by electric or magnetic fields and therefore
should accurately point to their sources\footnote{Since the $S^0$ is a neutral
spin-0 particle, even its magnetic dipole moment vanishes.}. That means that
if ultra-high-energy cosmic rays originate in persistent astrophysical objects
such as AGN's and if $S^0$'s are the primaries, the UHECR events will cluster
about certain directions in the sky.  It should also be possible to identify
a source ``behind" each UHE event.  In fact, four different "clusters" --
two pairs and two triplets -- have been identified among the highest energy
events.  The events in each of these clusters is consisent with pointing
directly to the same location.  A candidate astrophysical source, 3C 147,
has been identified for the triplet containing two of the highest energy
events.  At well over a gigaparsec, it is near enough for $S^0$'s to have
arrived with little energy loss.  

With the large sample of ultra-high-energy cosmic rays which hopefully
will be obtained by HiRes and the Auger Project, the question of clustering
will be settled.  Whether the UHECR events cluster or not, they may be due
to $S^0$'s.  If so, the prediction of a "GZK" cutoff in the spectrum, but
shifted to higher energy, can be tested.   

Laboratory experiments presently lack sufficient sensitivity to exclude
the possibility of a very light gluino, but that should change within a year
or two if the photino is light enough to account for dark matter.  




\end{document}